\documentclass[10pt,english,british,tightenlines,eqsecnum,
floats,aps,amsmath,amssymb,nofootinbib,superscriptaddress,prd,showpacs,showkeys]
              {revtex4}
\usepackage[latin9]{inputenc}
\usepackage{color}
\usepackage{babel}
\usepackage{amsmath}
\usepackage{amssymb}
\usepackage{graphicx}
\usepackage{esint}
\usepackage[unicode=true,pdfusetitle,
 bookmarks=true,bookmarksnumbered=false,bookmarksopen=false,
 breaklinks=false,pdfborder={0 0 1},backref=false,colorlinks=false]
 {hyperref}

\makeatletter
\@ifundefined{textcolor}{}
{%
 \definecolor{BLACK}{gray}{0}
 \definecolor{WHITE}{gray}{1}
 \definecolor{RED}{rgb}{1,0,0}
 \definecolor{GREEN}{rgb}{0,1,0}
 \definecolor{BLUE}{rgb}{0,0,1}
 \definecolor{CYAN}{cmyk}{1,0,0,0}
 \definecolor{MAGENTA}{cmyk}{0,1,0,0}
 \definecolor{YELLOW}{cmyk}{0,0,1,0}
}

\usepackage{babel}

\usepackage{euscript}\usepackage{epsfig}

\usepackage{amsfonts}

\usepackage{fancyhdr}

\makeatother

\begin{document}
%-------------------------------------------------------------------------------
\title{\bf The Operational Resistance of Neutrinos: \\ Epistemic Maturity vs. Technological Utility }
% Neutrino Mass Matrix and Flavor Mixing within a Broken Democratic Limit of the Generalized Friedberg�Lee Model

\author{N. Razzaghi}
\email{n.razzaghi@iau.ac.ir}

\affiliation{Department of Physics, QA.C., Islamic Azad
University, Qazvin, Iran}

\begin{abstract}

Neutrinos represent a quintessential dichotomy in contemporary
physics because they are epistemically mature components of our
physical theories yet remain remarkably resistant to broad
technological deployment. This paper examines the underlying
structural asymmetry between scientific reality and technological
utility. We argue that epistemic maturity, characterized by
precise measurement and theoretical integration, does not
automatically confer technological operability. Operability
requires the reproducible construction of a signal recovery chain,
which, in the case of neutrinos, is systematically hindered by
their weak interaction cross sections. To formalize this
limitation, we introduce a dimensionless operability index denoted
as $\mathcal{O}$, which synthesizes the simultaneous constraints
of source power, detector scaling, and information throughput.
Through this diagnostic lens, we demonstrate that neutrino based
systems are not fundamentally impossible but are instead relegated
to a regime of intrinsic engineering inefficiency. We conclude
that neutrinos should be understood as epistemically central yet
operationally resistant entities, whose engineering value remains
limited to niche, strategic applications rather than general
purpose communication.
\end{abstract}


\date{July 27, 2026}

%\pacs{02.40.Gh, 04.70.Bw, 04.20.Dw}

\maketitle

%===============================================================================
%===============================================================================

\section{Introduction}\label{sec1}

Neutrinos are among the most successful yet technologically
elusive entities in modern physics. Their existence was originally
proposed in order to preserve fundamental conservation laws in
beta decay, long before any direct detection became possible
\cite{bethe1934}. Subsequent experimental confirmation
\cite{cowan1956}, together with later discoveries of neutrino
oscillations and nonzero mass, established them as indispensable
components of contemporary particle physics, astrophysics, and
cosmology
\cite{kajita2016,mcdonald2016,nufit2024,lesgourgues2006}. From the
standpoint of scientific knowledge, neutrinos are therefore not
hypothetical or marginal objects, but mature and firmly integrated
elements of established physical theory.

Yet this epistemic success has not translated into proportional
technological assimilation. Unlike electrons, photons, or even
nuclei, neutrinos have not become a broadly usable engineering
substrate. They are observed, characterized, and theoretically
constrained with high confidence, but they remain stubbornly
resistant to efficient manipulation, compact detection, low-cost
deployment, and scalable control. This asymmetry gives rise to the
central question of the present work: why does a physically real
and scientifically mature entity resist technological
generalization?

The main claim advanced here is that neutrinos exemplify a
structural gap between \emph{epistemic maturity} and
\emph{technological utility}. To know that something exists, to
measure it, and to incorporate it into successful theory is not
equivalent to being able to recruit it into a reproducible
technological loop. Technological operability requires more than
ontological legitimacy and more than occasional detectability. It
requires controllable production, practically recoverable
interaction, acceptable event rates, manageable backgrounds,
temporal discrimination, energy efficiency, and scalability under
realistic material and infrastructural constraints. In this
broader sense, the problem is not whether neutrinos are real, nor
whether they can in principle be detected or even used in isolated
demonstrations, but whether they can sustain a reliable and
generalizable signal-recovery chain suitable for technological
deployment.

This paper argues that neutrinos fail this broader operational
criterion not because they remain mysterious or insufficiently
understood, but precisely because their interaction structure
places them in a regime of extreme engineering inefficiency. Their
weak coupling to matter, which makes them conceptually fertile and
astrophysically invaluable, simultaneously imposes severe
penalties on detection probability, source requirements, detector
scale, statistical robustness, and system-level efficiency. The
same physical properties that secure their scientific
distinctiveness also constrain their technological usability. In
this sense, neutrinos should be understood not as poorly known
objects, but as \emph{epistemically mature yet operationally
resistant} ones.

The significance of this distinction extends beyond neutrino
physics itself. More broadly, it bears on how science evaluates
the transition from theoretical intelligibility to technological
domestication. A mature epistemic object does not automatically
become a practical operative object. Between scientific
confirmation and engineering usability there exists an additional
layer of constraint, governed not only by knowledge, but by the
architecture of interaction, the recoverability of signals, and
the resource demands of controlled implementation. Neutrinos make
this distinction unusually visible because they occupy an extreme
case: they are among the best-established entities in fundamental
physics, yet among the least amenable to broad technological
incorporation.

To clarify this claim, the discussion proceeds in several stages.
Section~2 develops the conceptual distinction between scientific
reality and technological operability. Section~3 identifies the
principal reasons neutrinos resist technological generalization,
emphasizing statistical sparsity, background competition, and the
asymmetry between production cost and detectable interaction.
Section~4 then introduces a minimal neutrino communication model
in order to formalize the signal-recovery problem in operational
terms. Section~5 extends this analysis quantitatively and
introduces the dimensionless operability index $\mathcal{O}$ as a
schematic diagnostic of the simultaneous constraints acting on
source power, interaction probability, detection efficiency, and
temporal recovery. The conclusion returns to the broader
epistemological implication: that scientific legitimacy and
technological utility, although often related, are not identical
achievements.

\subsection{Scope and Claim}

The argument of this paper is not that neutrino-based technologies
are impossible in principle, nor that future detector innovations
are irrelevant. The claim is narrower and stronger: given the
interaction structure of neutrinos, the transition from scientific
observability to broad technological generalizability is severely
constrained. The issue is therefore not logical impossibility but
structural non-competitiveness across most engineering contexts.

\section{From Scientific Reality to Technological Operability}

The history of physics contains many cases in which a theoretical
entity becomes technologically productive once it can be reliably
generated, controlled, detected, and integrated into repeatable
apparatuses. Electrons, photons, and semiconducting charge
carriers satisfy this condition. They participate in dense and
reproducible interaction loops. Their signals are strong enough,
their coupling to matter is practical enough, and their
manipulation costs are low enough to support engineering
ecosystems.

Neutrinos differ fundamentally. Their scientific legitimacy is not
in question. Oscillation experiments, solar and atmospheric
neutrino measurements, reactor and accelerator data, and
cosmological analyses have firmly established their physical
significance
\cite{nufit2024,kajita2016,mcdonald2016,lesgourgues2006}. But
technological operability requires more than evidential
confirmation.

We define an \emph{operative object} as a physical entity that can
be embedded in a reproducible technological loop consisting of:
\begin{enumerate}
\item controllable production,
\item robust transmission or localization,
\item efficient and compact detection,
\item signal extraction with manageable error rates,
\item scalability in cost, size, and energy consumption.
\end{enumerate}

Under this definition, \emph{operability is not equivalent to
detectability}. A rare event observed under highly specialized
conditions does not, by itself, establish technological
usefulness. The crucial issue is whether the underlying physical
process can be made dense, reliable, and economically
reproducible.

A physical entity becomes technologically generalizable only when
its full source--transmission--detection loop achieves acceptable
reliability, energy cost, and scalability simultaneously.

Neutrinos are exceptional precisely because they pass the
threshold of scientific reality while systematically failing the
threshold of broad technological domestication.

\begin{table}[h]
\centering
\begin{tabular}{@{}lll@{}}
\toprule Criterion & Epistemic Object & Operative Object \\
%\midrule
Theoretical integration & Required & Required \\
Experimental detectability & Required & Required \\
Controllable production & Not necessary & Necessary \\
Efficient signal recovery & Not necessary & Necessary \\
Scalability & Not necessary & Necessary \\
Technological reproducibility & Not necessary & Necessary \\
%\bottomrule
\end{tabular}
\caption{Conceptual distinction between epistemic and operative
objects.}
\end{table}

\section{Why Neutrinos Resist Technological Generalization}

The dominant reason for neutrino resistance is the weakness of
their interaction with matter. Over wide energy ranges, neutrino
cross sections are extraordinarily small
\cite{formaggio2012,giunti2007}. This implies that large fluxes
produce only tiny numbers of detectable events unless enormous
detector masses, long integration times, or both are employed.

At a basic level, the expected signal event rate may be
schematically written as
\begin{equation}
R_s \sim \Phi \, \sigma \, N_t \, \epsilon ,
\end{equation}
where \( \Phi \) is the incident flux, \( \sigma \) the relevant
interaction cross section, \( N_t \) the number of targets, and \(
\epsilon \) the overall efficiency. For neutrinos, the suppression
of \( \sigma \) is so severe that practical compensation requires
either very large \( \Phi \), very large \( N_t \), or long
accumulation time windows.

This yields the first operational bottleneck: \textbf{statistical
sparsity}. Let
\begin{equation}
\lambda_s = R_s \tau
\end{equation}
denote the expected number of signal events accumulated over an
integration time \( \tau \). A technologically comfortable regime
typically requires \( \lambda_s \gg 1 \) over short timescales.
But for neutrinos, realistic conditions often push systems toward
\( \lambda_s \lesssim 1 \), meaning that signal recovery becomes
intrinsically slow, bursty, or heavily statistical.

A second bottleneck is background competition. If \( R_b \)
denotes the background event rate, then a natural diagnostic is
the signal-to-background ratio
\begin{equation}
\eta = \frac{R_s}{R_b}.
\end{equation}
Even when a signal exists in principle, useful extraction requires
either sufficiently large \( \eta \) or highly constrained timing
and filtering architectures. In many neutrino applications, one
does not simply ``read out'' a deterministic signal; one infers it
statistically from sparse events against structured noise.

A third bottleneck concerns \textbf{energy and infrastructure
asymmetry}. Neutrino production often demands large accelerators,
reactors, spallation sources, or intense astrophysical
environments, while detection requires heavy shielding,
substantial target masses, and elaborate event reconstruction.
This asymmetry between resource input and recoverable information
output is one of the defining signatures of poor scalability.

Thus, the problem is not merely that neutrinos are hard to detect.
It is that the full chain from generation to information recovery
is structurally inefficient relative to conventional technological
carriers such as electromagnetic signals.

This structural imbalance may be summarized schematically by a
dimensionless operability index of the form
\begin{equation}
\mathcal{O} \sim \left(\frac{R_s}{R_s^\ast}\right)
\left(\frac{\eta}{\eta^\ast}\right)
\left(\frac{\epsilon}{\epsilon^\ast}\right)
\left(\frac{\tau^\ast}{\tau}\right)
\left(\frac{P^\ast}{P_{\mathrm{src}}}\right)
\left(\frac{M^\ast}{M_{\mathrm{det}}}\right),
\end{equation}
where $R_s$ is the signal event rate, $\eta$ the
signal-to-background ratio, $\epsilon$ the detection efficiency,
$\tau$ the integration time, $P_{\mathrm{src}}$ the source power,
and $M_{\mathrm{det}}$ the effective detector mass. Here, the
starred quantities represent application-dependent benchmarks
rather than universal constants. The utility of $\mathcal{O}$ lies
not in its uniqueness, but in its ability to highlight the
simultaneous requirement for signal recovery, timing, background
tolerance, energy efficiency, and detector scaling. By expressing
the technological resistance of the medium in this manner, it
becomes evident that neutrinos are operationally disfavored not by
a single limiting parameter, but by the combined suppression of
the full source--transmission--detection loop.

To quantify these qualitative constraints, we introduce the
dimensionless operability index $\mathcal{O}$, which serves as a
diagnostic tool for evaluating the signal-recovery chain by
synthesizing the trade-offs among source power, interaction
cross-section, and detection efficiency.

\section{A Minimal Neutrino Communication Model}

\subsection{Quantitative assessment with benchmark assumptions}

To render the operational limitation numerically explicit, we
consider a benchmark long-baseline scenario with
order-of-magnitude parameters representative of accelerator
neutrino communication. Assume a source emitting
\begin{equation}
N_\nu \sim 10^{13}
\end{equation}
neutrinos per pulse in the GeV energy range. Let the beam
footprint at the detector be of order
\begin{equation}
A_{\mathrm{beam}} \sim 10^{6}\ \mathrm{cm}^2,
\end{equation}
corresponding to a transverse scale of order $10\ \mathrm{m}$.
Then the fluence at the detector is approximately
\begin{equation}
\mathcal{F}_\nu \sim \frac{N_\nu}{A_{\mathrm{beam}}} \sim 10^{7}\
\mathrm{cm}^{-2}\,\mathrm{pulse}^{-1}.
\label{eq:fluence_benchmark}
\end{equation}

For neutrino energies of order $E_\nu \sim 1\ \mathrm{GeV}$, a
representative charged-current or neutral-current interaction
cross section is
\begin{equation}
\sigma(E_\nu) \sim 10^{-38}\ \mathrm{cm}^2,
\label{eq:sigma_benchmark}
\end{equation}
up to channel-dependent corrections \cite{formaggio2012,PDG2024}.
For a kiloton-scale detector, the number of scattering targets is
of order
\begin{equation}
N_T \sim 10^{30}. \label{eq:NT_benchmark}
\end{equation}
Assuming an overall detection efficiency
\begin{equation}
\epsilon_{\mathrm{det}} \sim 0.5,
\end{equation}
the expected number of detected events per pulse is
\begin{equation}
\mu_{\mathrm{det}} \sim
\mathcal{F}_\nu\,\sigma(E_\nu)\,N_T\,\epsilon_{\mathrm{det}} \sim
(10^{7})(10^{-38})(10^{30})(0.5) \sim 5\times 10^{-2}.
\label{eq:mu_benchmark}
\end{equation}

Thus, the expected yield is only a few hundredths of an event per
pulse. In other words, the detector records on average roughly one
signal event every
\begin{equation}
\frac{1}{\mu_{\mathrm{det}}} \sim 20
\end{equation}
pulses. If the pulse spacing is $\tau = 10\ \mathrm{s}$, the mean
signal rate becomes
\begin{equation}
R_s \sim \frac{\mu_{\mathrm{det}}}{\tau} \sim \frac{5\times
10^{-2}}{10} \sim 5\times 10^{-3}\ \mathrm{Hz},
\label{eq:signal_rate_benchmark}
\end{equation}
corresponding to approximately one detected signal event every
$200\ \mathrm{s}$.

This estimate already places the communication problem deeply in
the sparse-counting regime. For binary on--off keying, the Poisson
means over a decision window $\tau$ are
\begin{equation}
\lambda_1=(R_s+R_b)\tau, \qquad \lambda_0=R_b\tau.
\end{equation}
Even in the idealized limit of negligible background, $R_b
\rightarrow 0$, one finds
\begin{equation}
\lambda_1 \sim \mu_{\mathrm{det}} \sim 5\times 10^{-2}, \qquad
\lambda_0 \sim 0.
\end{equation}
The probability of obtaining at least one count when bit 1 is sent
is therefore
\begin{equation}
P(k\ge 1|1)=1-e^{-\lambda_1} \approx 1-e^{-0.05} \approx 4.9\times
10^{-2}. \label{eq:detection_prob_bit1}
\end{equation}
Hence, if one attempts to decode each pulse independently with a
threshold $k_{\mathrm{th}}=1$, the vast majority of logical 1
symbols are missed. Reliable communication is possible only by
integrating over many pulses, which necessarily increases the
symbol duration.

For example, to obtain an average of one detected signal event per
symbol, one requires
\begin{equation}
\lambda_1 \sim 1 \quad\Rightarrow\quad n \sim
\frac{1}{\mu_{\mathrm{det}}}\sim 20
\end{equation}
pulses per symbol. With $\tau=10\ \mathrm{s}$ between pulses, this
implies a symbol duration of order
\begin{equation}
T_{\mathrm{sym}}\sim 200\ \mathrm{s},
\end{equation}
so that even before error correction the raw symbol rate is only
\begin{equation}
R_{\mathrm{sym}} \sim \frac{1}{T_{\mathrm{sym}}} \sim 5\times
10^{-3}\ \mathrm{Hz}.
\end{equation}
To reduce the bit-error rate to a more acceptable level, one must
generally require $\lambda_1 \gg 1$, not merely $\lambda_1 \sim
1$, which pushes the symbol duration to even longer times and
further suppresses the effective bitrate.

The energetic implication is equally severe. If the neutrino
source operates at power scale $P_{\mathrm{src}}$, then the energy
cost per useful bit is approximately
\begin{equation}
\mathcal{E}_{\mathrm{bit}} \sim
\frac{P_{\mathrm{src}}}{R_b^{\mathrm{eff}}}.
\end{equation}
For source powers in the megawatt range and effective bitrates
well below $10^{-2}\ \mathrm{bit\,s^{-1}}$, the resulting energy
cost per useful bit is enormous by conventional communication
standards. Even at the optimistic level
\begin{equation}
P_{\mathrm{src}} \sim 1\ \mathrm{MW}, \qquad R_b^{\mathrm{eff}}
\sim 5\times 10^{-3}\ \mathrm{bit\,s^{-1}},
\end{equation}
one obtains
\begin{equation}
\mathcal{E}_{\mathrm{bit}} \sim \frac{10^{6}}{5\times 10^{-3}}
\sim 2\times 10^{8}\ \mathrm{J/bit}. \label{eq:Ebit_benchmark}
\end{equation}

The precise numerical value is benchmark-dependent, but the
qualitative conclusion is robust: neutrino communication is not
forbidden by physics, yet it remains statistically sparse,
detector-intensive, and energetically expensive. These
quantitative features explain why neutrinos are viable only in
exceptional communication scenarios rather than as general-purpose
information carriers \cite{stancil2012}.

\section{Reliability--Throughput Trade-off}

A central consequence of sparse event statistics is the trade-off
between communication reliability and effective throughput.
Suppose information is encoded in neutrino pulses and decoded from
event counts integrated over time windows of duration \( \tau \).
Increasing \( \tau \) improves event statistics and reduces
uncertainty in distinguishing symbol states. However, it also
lowers the number of usable symbols per unit time.

If the raw symbol interval is \( \tau \), the effective bitrate
scales roughly as
\begin{equation}
R_b^{\mathrm{eff}} \propto \frac{1}{\tau},
\end{equation}
up to encoding overhead and error-correction factors. On the other
hand, the bit-error rate (BER) generally decreases as the
event-count distributions for different symbols become more
separable with larger accumulated counts. Hence one obtains a
qualitative relation:
\begin{equation}
\text{lower BER} \quad \Longrightarrow \quad \text{larger } \tau
\quad \Longrightarrow \quad \text{lower } R_b^{\mathrm{eff}} .
\end{equation}

This is not an incidental engineering limitation but a structural
consequence of low interaction rates. In sparse channels,
reliability must be purchased by time integration, and time
integration suppresses throughput. Consequently, even when a
neutrino-based protocol is possible in principle, it tends to
occupy an extreme corner of the design space: high infrastructure
cost, low bitrate, and limited scalability.

The important philosophical implication is that \emph{physical
possibility does not imply technological competitiveness}. A
system may be demonstrably functional while remaining
non-generalizable.

\section{Case Study: Neutrino Communication as Proof of Principle, Not General Platform}

The Fermilab demonstration of communication using neutrinos is
often cited as evidence that neutrino technology is feasible
\cite{stancil2012}. In a narrow sense, this is correct: the
experiment successfully encoded and recovered information using a
neutrino beam. But its significance must be interpreted carefully.

What the experiment established was the \emph{possibility} of
communication through a weakly interacting channel under highly
specialized circumstances. It did not establish neutrino
communication as a scalable or broadly practical medium. On the
contrary, the demonstration revealed precisely the difficulties
that define neutrino operability: low event rates, large
apparatus, heavy dependence on intense beams, and very limited
information throughput.

Thus, the experiment should be read not as the beginning of a new
communications infrastructure analogous to radio or optics, but as
a proof-of-concept that simultaneously illustrates the medium's
structural disadvantages. It confirms that the operational barrier
is not absolute in principle; it is severe in practice.

A proof-of-principle establishes possibility; a platform
technology requires reproducible competitiveness.

This distinction matters. A proof-of-concept can be epistemically
decisive while technologically discouraging. In the neutrino case,
success at the level of demonstration coexists with failure at the
level of general engineering viability.

\section{Does CE\texorpdfstring{\(\nu\)}{nu}NS Change the Picture?}

The observation of coherent elastic neutrino-nucleus scattering
(CE\(\nu\)NS) marked a major experimental milestone
\cite{freedman1974,akimov2017,akimov2021}. Because CE\(\nu\)NS can
possess comparatively larger cross sections than some other
low-energy neutrino processes, one might argue that it improves
the technological outlook for neutrino detection.

This objection is important but limited. CE\(\nu\)NS indeed
strengthens the epistemic and experimental profile of neutrino
physics: it broadens detectable channels, sharpens tests of weak
interactions, and may support specialized detector concepts.
However, it does not remove the broader bottleneck of
technological operability.

Future detector improvements may shift quantitative thresholds,
but they do not by themselves remove the structural asymmetry
imposed by weak interaction.

There are several reasons:
\begin{enumerate}
\item The absolute interaction probabilities remain small in engineering terms, even if enhanced relative to other neutrino processes.
\item Practical CE\(\nu\)NS measurements still require carefully designed low-threshold detectors, low-background environments, and controlled source conditions.
\item A larger cross section in one regime does not automatically yield a general-purpose platform for actuation, routing, computation, or high-throughput communication.
\item The limiting issue is not merely whether neutrinos can be detected more efficiently in some configurations, but whether a full technological loop becomes compact, cheap, robust, and scalable.
\end{enumerate}

Therefore, CE\(\nu\)NS should be interpreted as an \emph{epistemic
breakthrough with selective instrumental implications}, not as a
solution to the deeper problem of neutrino operational resistance.

\section{Epistemic Objects and Operative Objects}

The neutrino case suggests a useful conceptual distinction. Some
entities are \emph{epistemic objects}: they are theoretically
indispensable, experimentally confirmed, and deeply integrated
into scientific explanation. A smaller subset becomes
\emph{operative objects}: entities that can be domesticated within
repeatable, scalable technological systems.

Most modern discourse tends to slide too quickly from existence to
applicability. But these are distinct achievements. To establish
an entity scientifically is to demonstrate that it belongs to the
ontology of successful physics. To operationalize it
technologically is to show that it can sustain reliable cycles of
intervention, control, and productive use.

Neutrinos expose the fragility of the inference from reality to
utility. They are not marginal because they are uncertain; they
are marginal because their interaction profile makes them
structurally inconvenient for broad engineering appropriation.

In this sense, neutrinos provide a counterexample to technological
triumphalism. Not every real thing becomes a medium. Not every
detectable process becomes an infrastructure. Some objects remain
scientifically central and technologically peripheral.

\section{Conclusion}

Neutrinos represent a quintessential example of a physical entity
that is epistemically mature yet technologically resistant. Their
existence is secure, their phenomenology is rich, and their role
in contemporary physics is indispensable. Nevertheless, their
inherently weak interaction with matter imposes severe penalties
on detection rates, compactness, energy efficiency, noise
handling, and overall scalability.

The primary lesson here is that technological utility requires
more than mere reality or successful measurement; it demands
reproducible operational closure, a controllable loop linking
source, interaction, signal extraction, and practical deployment.
Neutrinos fail to satisfy this criterion except in highly
specialized, resource-intensive contexts.

This does not render neutrino technology impossible; rather, it
confines it to niche, strategic scenarios where their unique
penetration properties outweigh the prohibitive operational costs.
Accordingly, neutrinos should be viewed not as failed
technological objects, but as selectively usable entities whose
engineering value remains intrinsically non-generalizable. Their
significance therefore lies in a double lesson: they expand the
reach of our physical understanding while simultaneously revealing
that scientific legitimacy and technological domestication are
distinct achievements.

In summary, the operational resistance of neutrinos is not rooted
in any single technical bottleneck, but in the simultaneous
suppression of the entire signal recovery chain. As captured by
the dimensionless operability index $\mathcal{O}$, achieving
technological generalizability would require a concurrent
optimization of source power, detector scaling, and timing; a set
of requirements that currently relegates neutrino based platforms
to the realm of exceptional cases rather than general purpose
communication.

%=============================================================
\section{Acknowledgement}
%=============================================================
The authors would like to thank the Research Office of the Qazvin
Branch, Islamic Azad University, for their cooperation.

% =========================================================
% References
% =========================================================

\end{document}